\pdfoutput=1
\documentclass{article}
\usepackage{frascatiphys,here,graphicx,subfigure}

\usepackage{braket}
\usepackage{amsmath}
\usepackage{wrapfig}

\begin{document}
\title{\textbf{Status of measurement of $K_S \rightarrow \pi e \nu$ branching ratio and lepton
charge asymmetry with the KLOE detector }
}
\author{
Daria Kami\'nska      \\
on behalf of the KLOE-2 collaboration \\
{\em Institute of Physics, Jagiellonian University, Cracow, Poland}
}
\maketitle
\baselineskip=11.6pt
\begin{abstract}
 We present the current status of the analysis of about 1.7 billion $K_S K_L$ pair events collected at DAFNE 
 with the KLOE detector to determine the branching ratio of $K_S \rightarrow \pi e \nu$ decay and
 the lepton charge asymmetry. This sample is $\sim 4$ times larger in statistics
 than the one used in a previous KLOE analysis, allowing us to improve the accuracy
 of the measurement and of the related tests of $\mathcal{CPT}$ symmetry and $\Delta S= \Delta Q$ rule.   
\end{abstract}
 \baselineskip=14pt
\section{Introduction}
 A special role in $\mathcal{CPT}$ violation searches plays a neutral kaon
 system which, due to a sensitivity to a variety of symmetry violation effects, is one
 of the best candidates for such kind of studies. One of the possible tests is based on
 the comparison between semileptonic asymmetry in $K_S$ decays ($A_S$) and the analogous
 asymmetry in $K_L$ decays ($A_L$)\cite{handbook_cp}. So far the $A_L$\cite{ktev_kl_charge_asymm} was determined with a precision
 more than two orders of magnitude better than $A_S$\cite{kloe_final_semileptonic}. The present accuracy
 of $A_S$ determination is dominated by the statistical uncertainty. Therefore the  aim
 of this work is a determination of $A_S$ with two times smaller statistical error due to
 four times bigger data sample and improved systematical uncertainties.
 Charge asymmetry is also an important source of information about the real and imaginary parts of
 the $\mathcal{CPT}$ violating parameter~$\delta_{K}$. Until now $Re(\delta_K)$ was known with much
 worse precision than $Im(\delta_K)$. 

 The measurement was performed using KLOE detector\cite{kloe} located
 at DA$\Phi$NE accelerator\cite{dafne} in the National Laboratory in Frascati, Italy. The experimental data used
 in this paper has been collected during data campaign in 2004-2005.

\section{Charge asymmetry in neutral kaons semileptonic decays}
 Neutral kaons are the lightest particles which contain a strange quark. Observed short-living $K_S$
 and long-living $K_L$ are linear combinations of strange eigenstates ($K^{0}$ and~$\bar{K^{0}}$):
 \begin{equation}
  \begin{aligned}
    \ket{ K_{S} } & = \frac{1}{\sqrt{ 2(1+|\epsilon_{S}|^2) }} \left( (1+ \epsilon_{S})  \ket{K^0}  +
    (1- \epsilon_{S})  \ket{\bar{K^0}} \right),   \\ 
    \ket{ K_{L} } & = \frac{1}{\sqrt{2(1+|\epsilon_{L}|^2)}  } \left(  (1+ \epsilon_{L})  \ket{K^0}
    -(1- \epsilon_{L})  \ket{\bar{K^0}} \right).  
    \label{ks_kl}
  \end{aligned}
  \end{equation}
 where introduced small parameter $\epsilon_{S}$ and $\epsilon_{L}$ can be  rewritten to separate 
 $\mathcal{CP}$ and $\mathcal{CPT}$ violation parameters $\epsilon_{K}$ and $\delta_{K}$, respectively:
     \begin{align}
      \begin{aligned}
       \epsilon_{S} = \epsilon_{K} + \delta_{K}, \\
       \epsilon_{L} = \epsilon_{K} - \delta_{K}.  
      \end{aligned}
      \label{epsilon_and_symmetries}
     \end{align}
 \begin{figure}[h!]
   \centering
     \includegraphics[width=\textwidth]{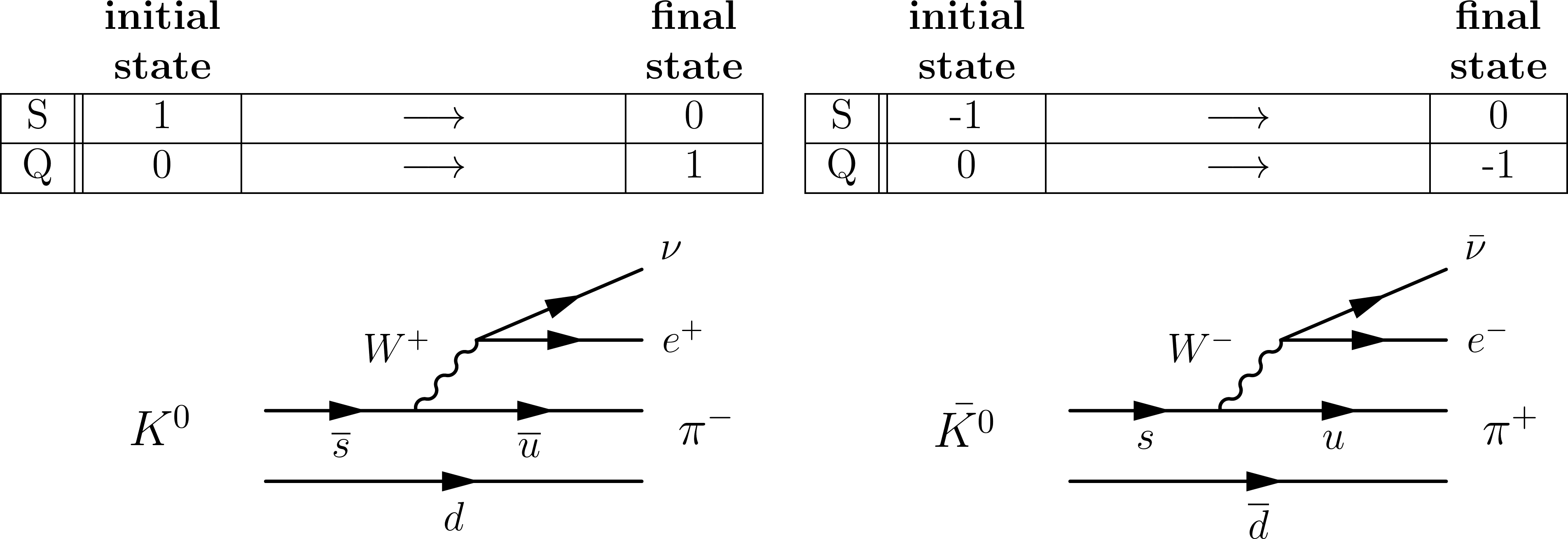}
   \caption{Feynman diagrams for $K^0$ and $\bar{K^0}$ semileptonic decay.} 
   \label{feynmann_diagrams_semi} 
  \end{figure}
 In the Standard Model decay of $K^0$ (or $\bar{K^0}$) state is associated with the
 transition of the $\bar{s}$ quark into $\bar{u}$ quark (or $s$ into $u$) and emission of the
 charged boson. 
 Change of strangeness ($\Delta S$) implies the corresponding
 change of electric charge  ($\Delta Q$) (see Figure~\ref{feynmann_diagrams_semi}). This is so called $\Delta S = \Delta Q$ rule.
 Therefore decays  of $K^0\rightarrow \pi^- e^+ \nu$ and $\bar{K^0} \rightarrow \pi^+ e^-
 \bar{\nu}$ are present but $ K^0 \rightarrow \pi^+ e^-\bar{\nu}$ and $\bar{K^0} \rightarrow \pi^- e^+
 \nu$ are not.
 Decay amplitudes for semileptonic decays of states $\ket{ K^0}$ and $\ket{ \bar{K^0}}$  can
 be written as~follows\cite{handbook_cp}:
    \begin{align}
    \begin{aligned}
     \label{aplitudes_def}
     \bra{\pi^{-} e^{+} \nu } H_{weak}  \ket{K^{0}}              & = a + b         \\ 
     \bra{\pi^{+} e^{-} \bar{\nu} } H_{weak}  \ket{ \bar{K^{0}}} & = a^{*} - b^{*} \\ 
     \bra{\pi^{+} e^{-} \bar{\nu} } H_{weak}  \ket{K^{0}}        & = c + d         \\ 
     \bra{\pi^{-} e^{+} \nu } H_{weak}  \ket{ \bar{K^{0}}}       & = c^{*} - d^{*} \\ 
    \end{aligned} 
    \end{align} 
 where the $H_{weak}$ is a part of Hamiltonian corresponding to the weak interaction and
 $a$,~$b$,~$c$,~$d$ parameters describe the semileptonic decay amplitudes. 
 Applying the symmetry operators to above amplitudes  a set of relations between them can be obtained (Table~\ref{tab_relations}).
 Semileptonic decay amplitudes can be associated with the  $K_{S}$
 and $K_{L}$ semileptonic decay widths through the charge asymmetry~($A_{S,L}$):
  \begin{align}
      \label{eq4} 
    A_{S,L} & =      
      \frac{\Gamma(K_{S,L} \rightarrow \pi^{-} e^{+} \nu) - \Gamma(K_{S,L}
       \rightarrow
       \pi^{+} e^{-}
       \bar{\nu})}{\Gamma(K_{S,L} \rightarrow \pi^{-} e^{+} \nu) + \Gamma(K_{S,L}
       \rightarrow
       \pi^{+} e^{-} \bar{\nu})}   \nonumber \\
         & =  2 \left[ Re \left( \epsilon_{K} \right) \pm Re \left( \delta_{K}  \right) 
              + Re \left( \frac{b}{a} \right) \mp Re \left( \frac{d^{*}}{a} \right) \right]
                \\    
         &   \mbox{ if } \Delta Q = \Delta S  \nonumber \\
         & = 2 \left[ Re \left( \epsilon_{K} \right) \pm Re \left( \delta_{K}  \right) 
            + Re \left( \frac{b}{a} \right) \right] \nonumber \\
         &  \mbox{ if } \mathcal{CPT}  \mbox{ and } \Delta Q = \Delta S  \nonumber \\
         & = 2 \left[ Re \left( \epsilon_{K} \right)  \right]    \nonumber 
    \end{align}
 where above equation contains only the first order of symmetry-violating terms.
 Moreover, conservation of $\Delta Q = \Delta S$ rule and $\mathcal{CPT}$ symmetry simplifies
 Equation~\ref{eq4}.
 Determination the value of charge asymmetry for $K_{S}$ and $K_{L}$ allows for tests the fundamental assumptions 
 of Standard Model.

  \begin{table}[h]
   \begin{center}
   \begin{tabular}{|c||c|c|c|c|} \hline
      & $\mathcal{CP}$                & $\mathcal{T}$     & $\mathcal{CPT}$   & $\Delta S = \Delta
      Q$ \\ \hline \hline
    a & \textit{Im} = 0 & \textit{Im} = 0 &       &  \\
    b & \textit{Re} = 0 & \textit{Im} = 0 &  = 0  &  \\
    c & \textit{Im} = 0 & \textit{Im} = 0 &       & = 0 \\
    d & \textit{Re} = 0 & \textit{Im} = 0 & = 0    & = 0 \\ \hline
   \end{tabular} 
   \caption{Relations between discrete symmetries and semileptonic amplitudes.} 
   \label{tab_relations}
   \vspace{-20pt}
   \end{center}
  \end{table}

 The charge asymmetry for $K_{L}$ decays was precisely determined by KTeV
 experiment at Fermilab\cite{ktev_kl_charge_asymm}. 
 The measurement was based on  1.9 millions
 $K_{L} \rightarrow \pi^{\pm} e^{\mp} \nu$ decays produced in collision of proton beam with BeO
 target.
 At present the most accurate measurement of $K_{S}$
 charge asymmetry was conducted by KLOE collaboration\cite{kloe_final_semileptonic}. 
 Obtained charge asymmetry  for $K_{S}$ decays is
 consistent in error limits with charge asymmetry for $K_{L}$ decays
  which suggest
 conservation of $\mathcal{CPT}$ symmetry. However, this result is dominated  by a statistical
 uncertainty which is three times larger than the systematic one. Nevertheless it can be improved by analysing  
 1.7 $\mbox{fb}^{-1}$ total luminosity data sample acquired in 2004 and 2005.

\section{Measurement}
 The KLOE experiment located at DA$\Phi$NE $\phi$ factory is specially suited for analysis of $K_S
 \rightarrow \pi e \nu$ decay.
 The KLOE detector is constituted by two main components: a drift chamber and an electromagnetic
 calorimeter, both inserted into an axial magnetic field (0.52~T).
 Due to the size of the KLOE drift chamber, about 40\% of $K_L$ mesons decay inside the detector
 while the rest reach
 the electromagnetic calorimeter. Detection of $K_{L}$ interaction in the calorimeter allows to 
 identify a $K_{S}$ meson on the opposite side of $\phi$ meson decay point. 

 In order to select semileptonic decay of
 $K_S$ meson, an additional kinematic selection is applied. It starts from a requirement
 of a vertex formed by tracks of two oppositely charged particles near the interaction point. 
 Those tracks must be associated to calorimeter clusters. Obtained tracks
 parameters allow for identification of charged particles in the final state by applying a
 Time of Flight technique.
 \begin{figure}
  \includegraphics[width=0.45\textwidth]{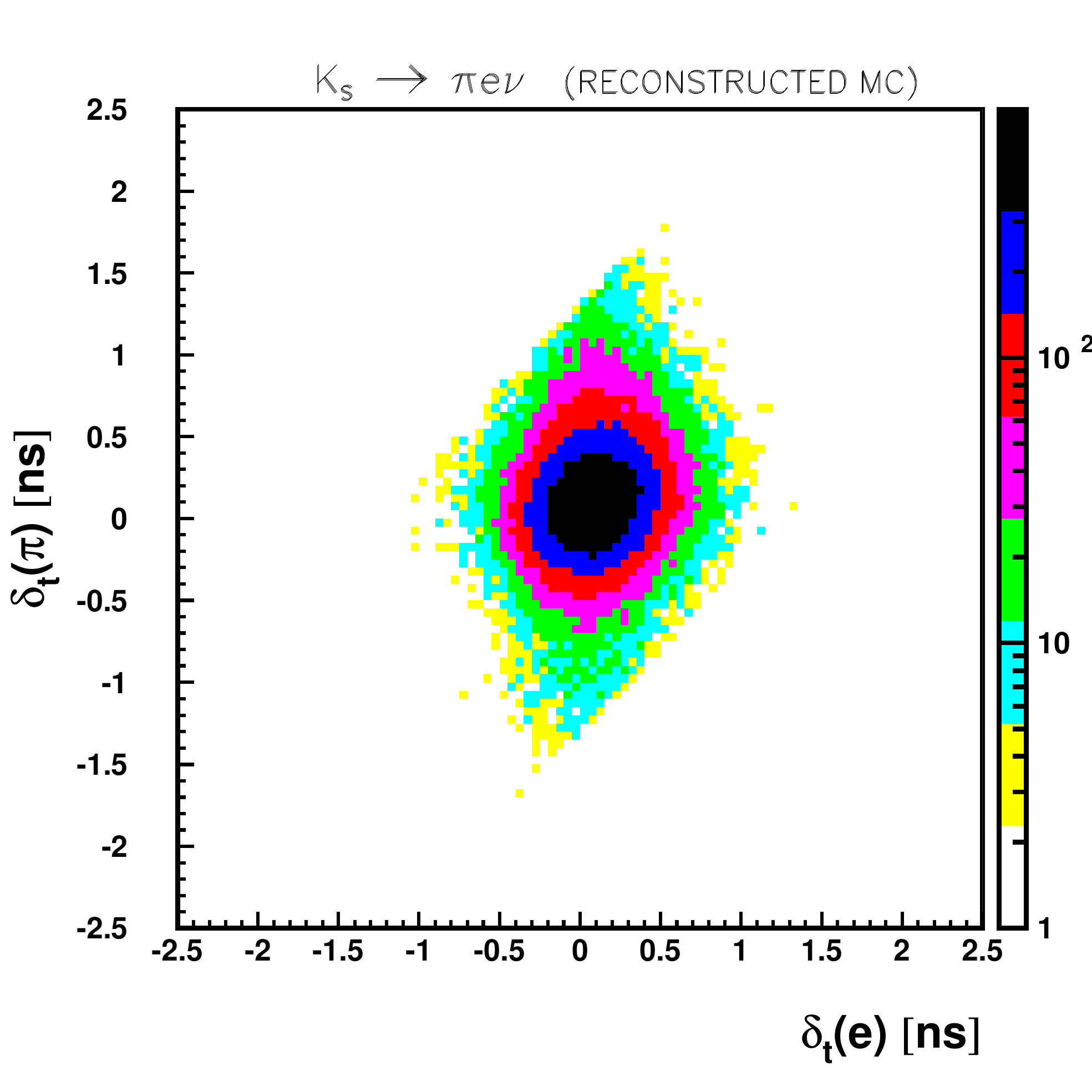}
  \includegraphics[width=0.45\textwidth]{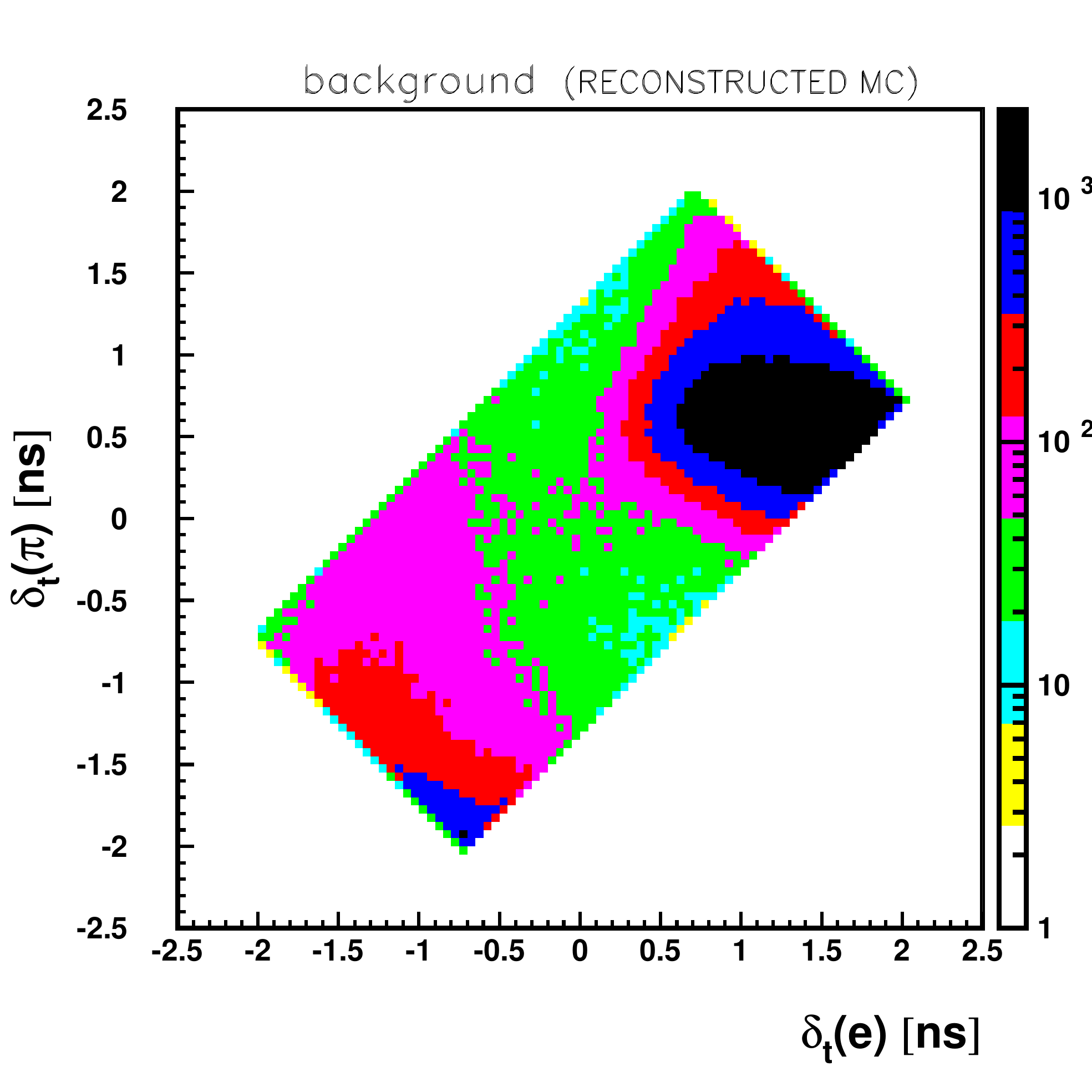}
  \caption{Distributions of the time difference for pion mass hypothesis ($\delta t(\pi)$) versus the
  time difference for electron mass hypothesis ($\delta t (e)$).
  Simulations of
  $K_S \rightarrow \pi e \nu$  and
  background events are shown on left and right plot, respectively.}
 \end{figure}
 \begin{wrapfigure}{r}{0.5\textwidth}
 \vspace{-20pt} 
 \includegraphics[width=0.45\textwidth]{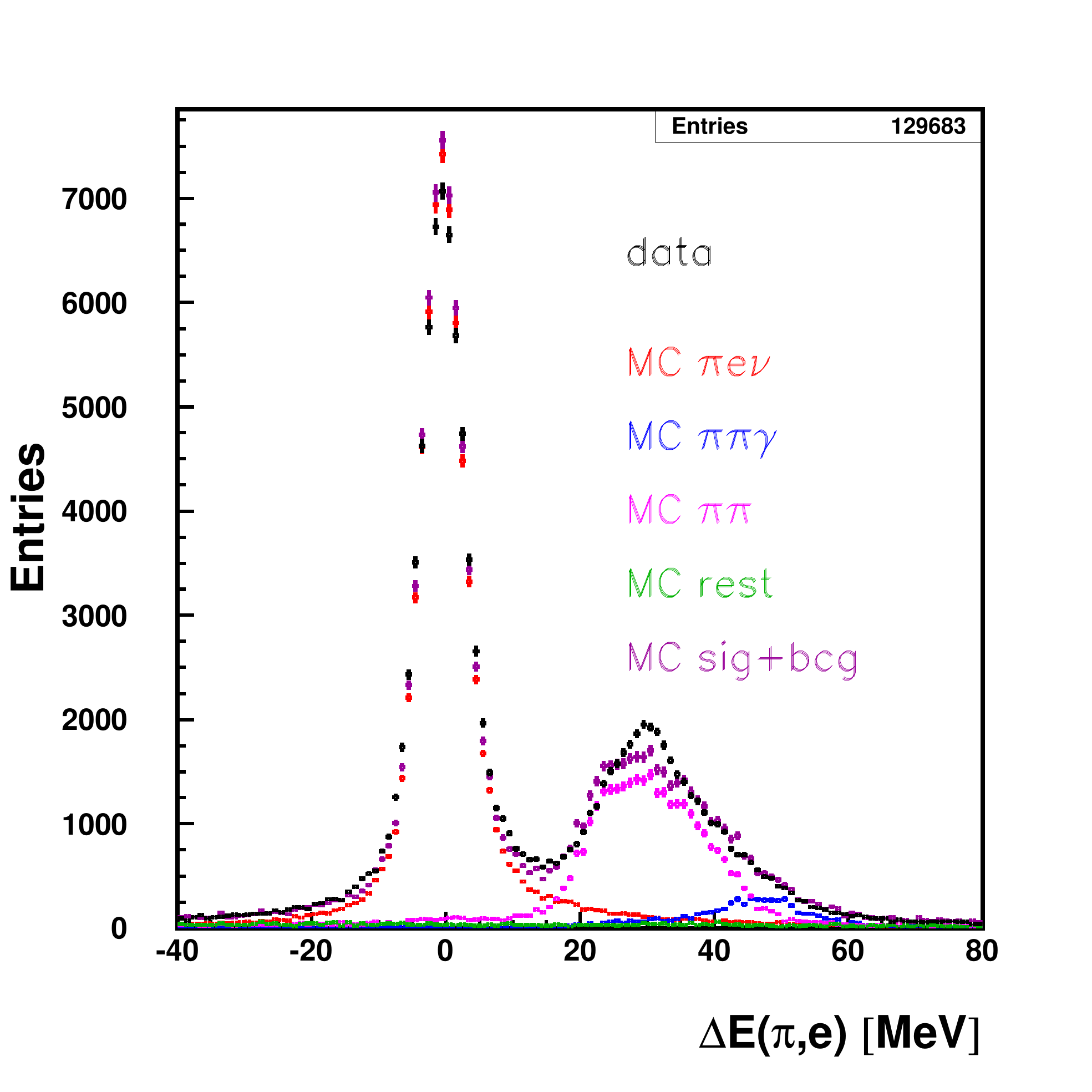}
 \caption{Distribution of $\Delta E(\pi,e) = E_{miss} - p_{miss}$ for all selected events 
 after normalization procedure. Meaning of simulated histograms is described in the legends.}
 \label{de_pi_e}
 \vspace{-30pt} 
 \end{wrapfigure}
 For each particle, the difference $\delta_{t}$ between the measured time of associated cluster 
 ($t_{cl}$) and time of flight is calculated assuming a given mass hypothesis,~$m_{x}$:
  \begin{equation}
   \begin{aligned}
     \delta_{t}(m_{x}) = t_{cl} - \frac{L}{c \cdot \beta(m_{x})},  \\
     \beta(m_{x}) =  \frac{P}{ \sqrt{P^2 + m_{x}^2 }},
   \end{aligned}
  \end{equation}
 where $L$ is a total length of particle trajectory  and $P$ is particle momentum measured by
 drift chamber. 
 The Time of Flight technique aims at rejection of the background, which is mainly
 due to $K_{S} \rightarrow \pi^+ \pi^-$ events, and  at identification of the final charge
 states  ($\pi^- e^+ \nu$ and $\pi^+ e^- \bar{\nu}$). The distribution of difference between
 missing energy and momentum ($\Delta E(\pi,e)$) shows remaining background components (see~Figure~\ref{de_pi_e}).
 Altogether around $10^5$ of $K_S \rightarrow \pi e \nu$ decays was
 reconstructed, which will be used to determine the charge asymmetry and branching
 ratio for $K_S$ semileptonic decays. The analysis is still in progress and preliminary results
 will be available soon\cite{my_master}.

\section*{Acknowledgments}
 This work was supported in part by the Foundation for
 Polish Science through the project HOMING PLUS BIS/2011-4/3 and the MPD programme;
 by~the Polish National Science Centre through the Grants No. 0469/B/H03/2009/37,
 0309/B/H03/2011/40, 2011/03/N/ST2/02641,
 2011/01/D/ST2/00748, 2011/ 03/N/ST2/02652, 2013/08/M/ST2/00323 and by 
 the EU Integrated Infrastructure Initiative Hadron Physics
 Project under contract number 
 RII3-CT-2004-506078; by~the European Commission under the 7th Framework Programme through the
 'Research Infrastructures'
 action of the 'Capacities' Programme, Call: FP7-INFRASTRUCTURES-2008-1, Grant Agreement
 No.~$227431$.

\end{document}